\documentclass[prd,preprint,superscriptaddress]{revtex4}
\usepackage{graphicx}
\usepackage[centertags]{amsmath}
\usepackage{amsfonts}
\usepackage{amssymb}
\usepackage{amsthm}
\usepackage{newlfont}
\usepackage{float}
\usepackage{longtable}
\usepackage[english]{babel}
\newcommand{\ds}{\displaystyle}
\newcommand{\df}{\displaystyle\frac}
\begin{document}
\selectlanguage{english}
\title{Observational constraints on interacting quintessence models}
\author{Germ\'{an} Olivares\footnote{E-mail address: german.olivares@uab.es}}
\affiliation{Departamento de F\'{\i}sica, Universidad Aut\'{o}noma
de Barcelona, Spain}
\author{Fernando Atrio-Barandela\footnote{E-mail address: atrio@usal.es}}
\affiliation{Departamento de F\'{\i}sica Te\'orica, Universidad de
Salamanca, Spain}
\author{Diego Pav\'{o}n\footnote{E-mail address: diego.pavon@uab.es}}
\affiliation{Departamento de F\'{\i}sica, Universidad Aut\'{o}noma de
Barcelona, Spain}
\begin{abstract}
We determine the range of parameter space of a Interacting
Quintessence Model that best fits the recent WMAP measurements of
Cosmic Microwave Background temperature anisotropies. We only
consider cosmological models with zero spatial curvature. We show
that if the quintessence scalar fields decays into cold dark
matter at a rate that brings the ratio of matter to dark energy
constant at late times, the cosmological parameters required to
fit the CMB data are: dark energy density $\Omega_{x} = 0.43\pm
0.12$, baryon fraction $\Omega_{b}=0.08\pm 0.01$, slope of the
matter power spectrum at large scales $n_{s}=0.98\pm 0.02$ and
Hubble constant $H_{0}=56\pm 4\,km/s/Mpc$. The data prefers a dark
energy component with a dimensionless decay rate parameter $c^{2}=
0.005$ and non-interacting models are consistent with the data
only at the 99.9\% confidence level. Using the Bayesian
Information Criteria we show that this extra parameter fits the
data better than models with no interaction. The quintessence
equation of state parameter is less constrained; i.e., the data
sets an upper limit $w_{x}\le -0.86$ at the same level of
significance. When the WMAP anisotropy data are combined with
supernovae data, the density parameter of dark energy increases to
$\Omega_{x}\simeq 0.68$ while $c^2$ augments to $6.3 \times
10^{-3}$. Models with quintessence decaying into dark matter
provide a clean explanation for the coincidence problem and are a
viable cosmological model, compatible with observations of the
CMB, with testable predictions. Accurate measurements of baryon
fraction and/or of matter density independent of the CMB data,
would support/disprove these models.
\end{abstract}
\pacs{98.80.Es, 98.80.Bp, 98.80.Jk}
\maketitle

\section{Introduction}
The recent observations of high redshift supernovae \cite{hubble},
Cosmic Microwave Background (CMB) temperature anisotropies
\cite{wmap} and the shape of the matter power spectrum \cite{sdss}
consistently support the idea that our  Universe is currently
undergoing an epoch of accelerated expansion \cite{reviews}.
Currently, the debate is centered on when did the acceleration
actually start and what agent is driving it. A variety of models
based on at least two matter components (baryonic and dark) and
one  dark energy component (with negative pressure) have been
suggested -see \cite{peebles_rmph}. The $\Lambda$CDM model, where
a vacuum energy density or cosmological constant provides the
negative pressure, was the earliest and simplest to be analyzed.
While this model is consistent with the observational data (high
redshift supernova \cite{hubble}, CMB anisotropies
\cite{wmap,boomerang}, galaxy cluster evolution \cite{sdss}), at
the fundamental level it fails to be convincing: the vacuum energy
density falls below the value predicted  by any sensible quantum
field theory by many orders of magnitude \cite{weinberg}, and it
unavoidably leads to the {\em coincidence problem}, i.e., ``Why
are the vacuum and matter energy densities of precisely the same
order today?" \cite{coincidence}. More sophisticated models
replace $\Lambda$ by a dynamical dark energy either in the form of
a scalar field (quintessence), tachyon field, phantom field or
Chaplygin gas. These models fit the observational data but it is
doubtful that they solve the coincidence problem \cite{lad}.

Recently, it has been proposed that dark matter and dark energy
are coupled and do not evolve separately
\cite{amendola,amendola0,iqm0,iqm,peebles,hoffman,mangano,rong}.
In particular the Interacting Quintessence (IQ) models of
references \cite{amendola,amendola0,iqm0,iqm}, aside from fitting
rather well the high redshift supernovae data, quite naturally
solve the coincidence problem by requiring the ratio of matter and
dark energy densities to be constant at late times. The coupling
between matter and quintessence is either motivated by high energy
particle physics considerations \cite{amendola} or is constructed
by requiring the final matter to dark energy ratio to be stable
against perturbations \cite{iqm, iqm0}. Since the nature of dark
matter and dark energy are unknown there are no physical arguments
to exclude their interaction. On the contrary, arguments in favor
of such interaction have been suggested \cite{peebles}. As a
result of the interaction, the matter density drops with the scale
factor $a(t)$ more slowly than $a^{-3}$.

A slower matter density evolution fits the supernovae data as well
as the $\Lambda$CDM concordance model does \cite{iqm}. The
interaction also alters the age of the Universe, the evolution of
matter and radiation perturbations and gives rise to a different
matter and radiation power spectra. All these effects will be used
to set constraints on the decay rate of the scalar field using
cosmological observations. In this paper, we shall further
constrain the Chimento et al. \cite{iqm} model by using the
recently WMAP measurements of the cosmic microwave background
temperature anisotropies. As it turns out, a small but
non-vanishing interaction between dark matter and dark energy is
compatible with the WMAP data with the advantage of solving the
coincidence problem. To some extent, this was already suggested in
a recent analysis that uses the position of the peaks and troughs
of the CMB \cite{peaks} to constrain a general class of
interacting models designed not to strictly solve the coincidence
problem, but to alleviate it \cite{scaling}. Briefly, the outline
of the paper is: in Section II we summarize the cosmological
model, in Section III we derive the equations of dark  matter and
dark energy density perturbations and find the range of parameter
space that best fits the observations; finally, in Section IV we
discuss our main results and present our conclusions.

\section{The interacting quintessence model}
The IQ model considered here has been constructed to solve the coincidence
problem by
introducing a coupling between matter and dark energy; their respective
energy densities
don't evolve independently. In this paper, we shall
simplify the Chimento {\em et al.} model \cite{iqm} in the sense
that the IQ will be assumed to decay into cold dark matter (CDM) and not into
baryons, as required by the constraints imposed by local gravity measurements
\cite{gravity, peebles_rmph}
The baryon--photon fluid evolves independently of the CDM
and quintessence components. Unlike \cite{iqm}, we do not include
dissipative effects.
In \cite{iqm} this scaling was considered only during matter domination,
so the
scalar field would evolve independently of the CDM component until it started
to decay at some early time. These assumptions facilitate the numerical work
while they preserve its essential features. Specifically,
the quintessence field (denoted
by a subscript $x$) decays into pressureless CDM (subscript $c$) according to
\cite{iqm}
\\
\begin{equation}\label{cont}
\begin{array}{rcl}
\displaystyle \frac{d\rho_c}{dt}+ 3H \rho_{c} = 3H c^{2}
\left(\rho_{c} + \rho_{x}\right),\\
\displaystyle \frac{d\rho_x}{dt} + 3(1+w_{x})H\rho_{x} = -3H c^{2}
\left(\rho_{c} + \rho_{x}\right),
\end{array}
\end{equation}
\\
where $w_{x} <0 $ is the equation of state parameter of the dark
energy and $c^{2}$ is a small dimensionless constant parameter
that measures the intensity of the interaction. Approaches similar
(but not identical) to ours have been discussed in
\cite{amendola,amendola0,hoffman,mangano,rong}. Eqs. (\ref{cont})
were not derived assuming some particle physics model for the
interaction, where quintessence is described as an scalar field
with a given potential. We followed a phenomenological approach
and instead we have required the Interacting Quintessence Model to
solve the coincidence problem. We have imposed the dark matter to
dark energy density interaction to give a dark matter to dark
energy ratio constant at late times and stable against
perturbations. As a result, the shape of the scalar field
potential is also fixed.

\begin{figure}
\centering
\includegraphics[scale=.8]{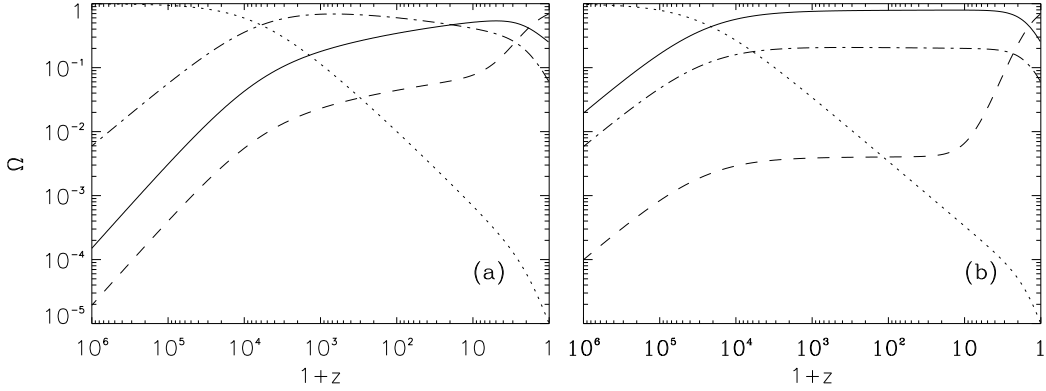}
\caption{Redshift evolution of different energy densities. Solid,
dashed, dotted and dot-dashed lines correspond to $\Omega_{c}$,
$\Omega_x$, $\Omega_{r}$ and $\Omega_{b}$, respectively. In panel
(a) $c^2=0.1$, and in panel (b)  $c^2=5\times 10^{-3}$.
The following parameters were assumed: $\Omega_{c,0}= 0.25$,
$\Omega_{x,0}= 0.7$,
$\Omega_{b,0}= 0.05$, $\Omega_{r,0}= 10^{-5}$, and $w_{x} = -0.99$.}
\label{fig:density}
\end{figure}

\begin{figure}\centering
\includegraphics[scale=.7]{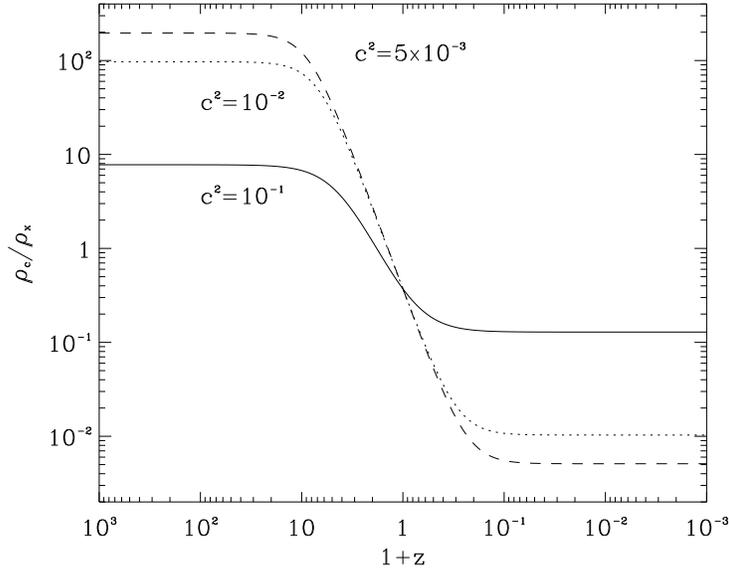}
\caption{Evolution of the ratio $r = \rho_{c}/\rho_{x}$ from
an unstable maximum
toward a stable minimum (at late times) for different values of $c^{2}$.
We took $r_{0}= 0.42$ as the current value.}
\label{fig:ratio}
\end{figure}

Eqs. (\ref{cont}) can be solved by Laplace transforming the system.
The result is
\\
\begin{equation}\label{density}
\begin{array}{lll}
\rho_{x}(a)&=&\df{H_0^2}{8\pi G w_{eff}}\ds
[3(c^2\Omega_{c,0}-(1-c^2)\Omega_{x,0})(a^{S_{+}}-a^{S_{-}})+
\Omega_{x,0}(S_{-}a^{S_{-}}
-S_{+}a^{S_{+}})],\\
\rho_{c}(a)&=&\df{H_0^2}{8\pi G w_{eff}}\ds
[3((1+w_x+c^2)\Omega_{c,0}+c^2\Omega_{x,0})(a^{S_{-}}-a^{S_{+}})+
\Omega_{c,0}(S_{-}a^{S_{-}}-S_{+}a^{S_{+}})],
\end{array}
\end{equation}
\\
where $w_{eff}=(w_{x}^{2}+4c^{2}w_{x})^{1/2}$, and
$S_{\pm}=-3(1+w_x/2) \mp (3/2) w_{eff}$. The density parameters
$\Omega_{c,0}$ and $\Omega_{x,0}$ denote the current values of
matter and dark energy, respectively. Solutions of Eqs.
(\ref{cont}) are plotted in Fig. \ref{fig:density}. Solid, dashed,
dotted and dot-dashed lines correspond to $\Omega_{c}$,
$\Omega_{x}$, $\Omega_{r}$ and $\Omega_{b}$, respectively. In
panel (a) $c^{2} = 0.1$ and there is a short period of baryon
dominance; this does not happen in panel (b) where  $c^{2} = 5
\times 10^{-3}$.
\\
\\
As detailed in Ref. \cite{iqm}, the
interaction dark matter--dark energy brings the
ratio $r \equiv \rho_{c}/\rho_{x}$ to a constant,
stable value at late times. From Eqs.(\ref{cont}),
it is seen that the evolution of the aforesaid ratio is
\\
\begin{equation}\label{ratio}
\frac{dr}{dt}=3H c^2\left[ r^{2}+\left(\frac{w_{x}}{c^{2}}+2\right)r+1\right].
\end{equation}
\\
The equation $dr/dt = 0$, has two stationary solutions, namely,
\[
r_{\pm}= -w_{x}/(2c^{2})-1\pm [w_x^{2}/(4c^{4})+w_x/c^{2}]^{1/2}\, ,
\]
\\
which verify $r_{+}\, r_{-}=1$ (with $r_{+} > r_{-}$). As shown in
Fig. \ref{fig:ratio}, the ratio evolves from an unstable maximum
$r_{+}$ at early times -with dark matter and quintessence energy
densities scaling as $a^{S_{+}}$- to a stable minimum
$r_{-}$ at late times, where both energy densities scale
as $a^{S_-}$. In Fig. \ref{fig:ratio}, the smaller the
coupling constant the larger the
ratio of cold dark matter to dark energy in the past and the smaller in
the future without significantly affecting the length of the transition period.
We are not suggesting that the Universe
is already in the late time epoch of constant, stable ratio
$r_{-}$. The value of the asymptotic ratio is
determined by the strength of the interaction and at present
this ratio could be still slowly evolving in time.

In terms of a scalar field description, the second equation of (\ref{cont})
is equivalent to \\
\begin{equation}\label{KG}
\frac{d^2\phi}{dt^2}
+ 3H\frac{d\phi}{dt} + V^{\prime}_{eff}=0 \, ,
\end{equation}
where $\phi$ denotes the dark energy field and $V_{eff}(\phi)$ is the
effective potential. The latter is given by
\\
\begin{equation}\label{potential}
V^{\prime}_{eff}=\frac{dV(\phi)}{d\phi}+3H c^{2}
\left(\rho_{c} + \rho_{x}\right)/(d\phi/dt).\\
\end{equation}
If $r=$ constant, the potential has two asymptotic limits:
$V\propto e^{(-\phi)}$ during both matter domination and the
period of accelerated expansion, and $V\propto \phi^{-\alpha}$
with $\alpha > 0$ well within the radiation dominated period. A
detailed study has shown that only potentials that are themselves
power laws -with positive or negative powers- or exponentials of
the scalar field yield energy densities evolving as power laws of
the scale factor \cite{liddle_scherrer}. Potentials with
exponential and power-law behavior have been considered
extensively in the high energy physics literature. Exponential
potentials arise as a consequence of Kaluza--Klein type
compactifications of string theory and in $N=2$ supergravity while
inverse power law models arise in SUSY QCD (see
\cite{iqm0,copeland}). Potentials showing both asymptotic
behaviors have also been studied \cite{sahni_wang}, but at present
there are not satisfactory particle physics model to justify the
shapes of potentials of this type \cite{riazuelo}.

\begin{figure}\centering
\includegraphics[scale=1]{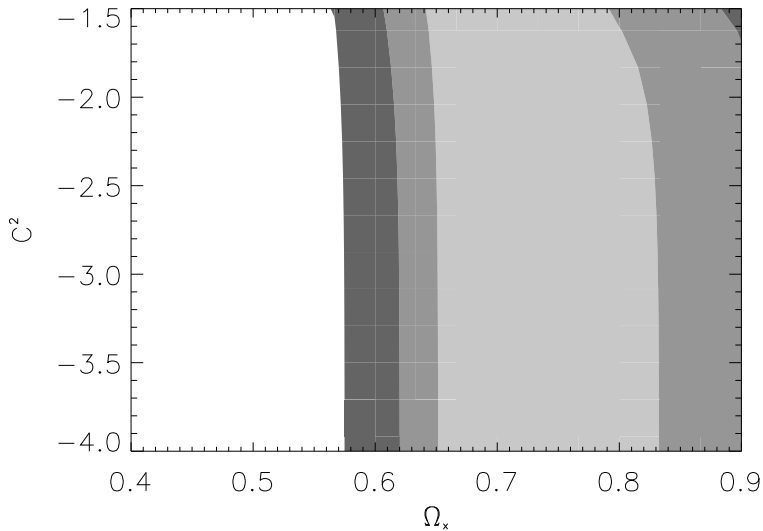}
\caption{Joint confidence intervals at 68\%, 95\% and 99.9\%
confidence level of IQM fitted to the ``gold" sample of SNIa data
of Riess {\it et al.} \cite{hubble}} \label{fig:likeSNIQM}
\end{figure}

 From the evolution of the background energy densities it is possible
to constrain the amplitude of the IQ and CDM coupling. Since
$w_{eff}> 0$, $c^2$ is confined to the interval $0\leq c^2<
|w_x|/4$. Negative values of $c^2$ would correspond to a transfer
of energy from the matter to the quintessence field and might
violate the second law of thermodynamics. Further constraints can
be derived by imposing stability of the interaction to first order
loop corrections \cite{one-loop}. In Fig. \ref{fig:likeSNIQM} we
used the supernova data of Riess {\it et al.} \cite{hubble} to
constrain model parameters. In the figure we plot the  68\%, 95\%
and 99.9\% confidence levels of a cosmological model after
marginalizing over, $w_x$ and the absolute magnitude of SNIa. We
set a prior: $-1.0\le w_x\le -0.6$. Variations of the baryon
density produce no significant differences and the Hubble constant
is unconstrained by this Hubble test since the absolute luminosity
of Type Ia supernovae is not accurately measured. The contours are
rather parallel to the $c^2$ axis, i.e., the low redshift
evolution of interacting models is not very different from the
non-interacting ones.

\section{Observational constraints on the matter-quintessence coupling}
Primordial nucleosynthesis and Cosmic Microwave Background temperature
anisotropies provide the best available tools
to constrain the physics of the early Universe. By
assumption, the scalar field decays into dark matter and not
into baryons. Since dark matter and quintessence density perturbations
are coupled to baryon and photons
only through gravity, there is no transfer of energy or momentum
from the scalar field to baryons or radiation. The evolution of density
perturbations of dark matter and dark energy can be simply derived from the
energy conservation equation. In the equations below we shall
use the conventions of \cite{bertschinger}. In the synchronous gauge,
\\
\begin{equation}\label{pert_eq_m}
\begin{array}{lll}
\dot\delta_{c}&=&-\df{\dot h}{2}-3\df{\dot a}{a}c^2\ds
      \left(\df{\delta_x}{r}+\delta_{c}\right),\\
\dot\theta_{c}&=&0,
\end{array}
\end{equation}
\\
while the evolution of dark energy density perturbations is given by
\begin{equation}\label{pert_eq_de}
\begin{array}{lll}
\dot\delta_{x}&=&-\ds(1+w_x)(\theta_x+\df{\dot h}{2})
-3\df{\dot a}{a}(c_s^2-w_x)\delta_x\\
&-&9\left(\df{\dot a}{a}\right)^2 \ds(c^2_{s,x}-w_x)(1+w_x)\theta_xk^{-2}
+3\df{\dot a}{a}c^2\ds(\delta_x+r\delta_{c})\\
\dot\theta_x  &=&-\ds(1-3c^2_{s,x})\df{\dot a}{a}\theta_x
+\df{k^2c^2_{s,x}}{1+w_x}\delta_x -3\df{\dot a}{a}\df{c^2}{1+w_x}(1+r)\theta_x,
\end{array}
\end{equation}
\\
where $\delta$ and $\theta$ denote  the density contrast and the
divergence of the peculiar velocity field of each component,
respectively, derivatives are with respect to conformal time and
$c^2_{s,x}$ is the quintessence sound speed, taken to be unity as
for a scalar field with a canonical Lagrangian. The interaction
introduces the terms with a $c^2$ factor on the right hand side of
Eqs. (\ref{pert_eq_m}) and (\ref{pert_eq_de}). Combining these
equations, the  evolution of density perturbations in the IQ field
are described by a driven damped harmonic oscillator, where the
driving term is the gravitational field \cite{caldwell}. After a
brief transient period, the evolution is dominated by the
inhomogeneous solution and is insensitive to the initial
amplitude. To find the model that best fits the WMAP data, we have
implemented equations (\ref{density}), (\ref{pert_eq_m}) and
(\ref{pert_eq_de}) into the CMBFAST code \cite{cmbfast}. We used
the likelihood code provided by the WMAP team \cite{likeli} to
determine the quality of the fit of every model to the data. Since
we are introducing a new parameter, the coupling between dark
matter and dark energy, the parameter space could become
degenerate with different local maxima representing models that
fit the data equally well. For this reason, we did not use a Monte
Carlo Markov Chain approach \cite{likeli} but we run through a
grid of models on a six-dimensional parameter space. Grids of
models are computationally very expensive. To make the
computations feasible we reduced the parameter space by
introducing prior information. We imposed two constraints: (1) all
models were within the 90\% confidence level of the constraint
imposed by Big Bang Nucleosynthesis: $0.017\le\Omega_bh^2\le
0.027$ \cite{olive} and (2) in all cosmologies the age of the
Universe was chosen to be $t_{0}>12$ Gyr. With these requirements,
we explore the region of parameter space close to the concordance
model. We have considered only flat models  with no reionization,
no gravitational waves and no running of the spectral index. We
considered a 6-dimensional parameter space and assumed our
parameters to be uniformly distributed in the following intervals:
Hubble constant $H_0=[46,90]\,km/s/Mpc$, baryon fraction $\Omega_b
=[0.01,0.12]$, dark energy $\Omega_x=[0.1,0.9]$, slope of the
matter power spectrum on large scales $n_s=[0.95,1.04]$, dark
energy equation of state $w_x=[-1.0,-0.65]$ and $c^2=[0,0.05]$. We
took 23, 15, 33, 10, 9 linear subdivisions and 22 logarithmic
subdivisions of the above intervals, respectively. The likelihood
was computed using the routines made publicly available by the
WMAP team.

\begin{figure}\centering
\includegraphics[scale=1]{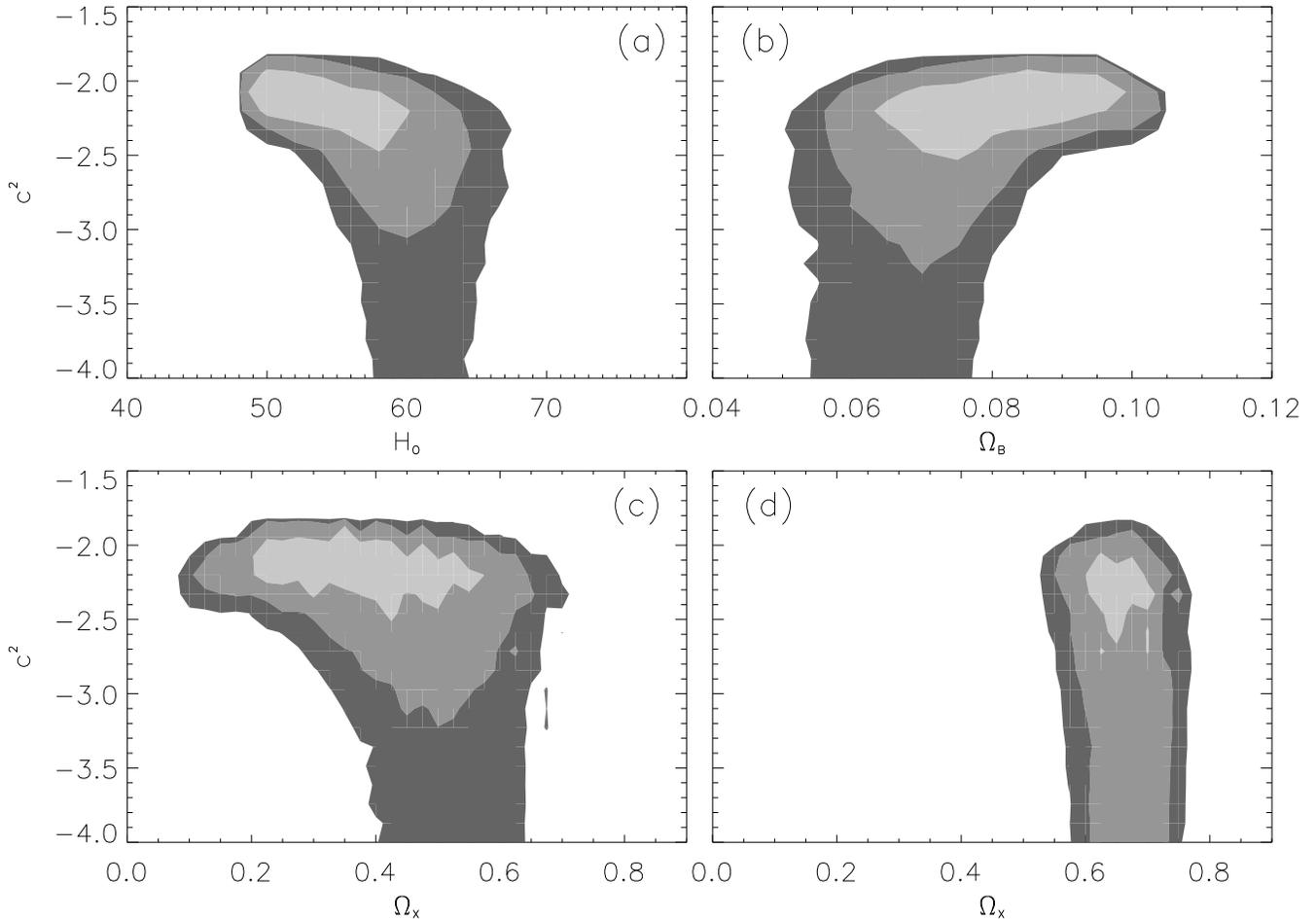}
\caption{ Joint confidence intervals at the 68\%, 95\% and 99.9\%
level for pairs of parameters after marginalizing over the rest.
For convenience the $c^2$ axis is represented using a logarithmic
scale and it has been cut to $c^2\le 10^{-4}$, though models with
$c^2 = 0$ have been included in the analysis. In panels (a), (b)
and (c) models were fit to CMB data alone. In panel (d) we
included supernovae data of Riess {\it et al.} \cite{hubble}.}
\label{fig:contour}
\end{figure}

In Fig.\ref{fig:contour} we give confidence intervals for
pairs of parameters after marginalizing over the rest.
Contours represent the 68\%, 95\%
and 99.9\% confidence levels.
Models with $c^2=0$ have been computed and were included in the
analysis.  The results were undistinguishable from those of $c^2 =
10^{-4}$ the last point included in the graphs. The WMAP data sets
strong upper limits on the quintessence decay rate. A non-zero
decay rate is clearly favored by the data. At $c^2\sim 10^{-2}$
the contours indicate a steep gradient in the direction of growing
$c^2$. This behavior is associated with the decreasing fraction of
CDM at recombination with increasing $c^2$.  When the interaction
rate is large, the Universe goes through a period dynamically
dominated by baryons (Fig.\ref{fig:density}a). The oscillations on
the baryon-photon plasma induce large anisotropies in the
radiation and those models are strongly disfavored by the data. In
Fig.\ref{fig:contour}, the models fit the data more comfortably
with lower values of $\Omega_x$ and $H_{0}$ than in the
$\Lambda$CDM concordance model. Interacting models have larger
dark energy density in the past than non-interacting models,
achieving the same rate of accelerated expansion today with  a
smaller $\Omega_{x,0}$. Our best model also requires larger baryon
fraction since the matter density is smaller prior to
recombination than in the concordance model, therefore dark matter
potential wells are shallower and a higher baryon fraction is
required to reproduce the amplitude of the first acoustic peak
\cite{cmb}. The mean value of the cosmological parameters and
their corresponding $1\sigma$ confidence intervals are:
$\Omega_{x} = 0.43\pm 0.12$, $\Omega_{b}=0.08\pm 0.01$,
$n_s=0.98\pm0.2$ and $H_{0}= 56\pm 4 \,km/s/Mpc$. The latter
number is not very meaningful since the probability distribution
of $H_0$ is rather skewed. As it can be seen in
Fig.\ref{fig:contour}a, low values of $H_0$  are suppressed very
fast. As the height of the first acoustic peak scales with
$\Omega_bh^2$, high values of baryon fraction are speedily
suppressed by the WMAP data, which translates into an even faster
suppression of low values of $H_0$. With respect the quintessence
equation of state, as we did not explore models with $w_{x}<-1$,
we can only set an upper limit $w_{x}\le -0.86$ at the 1$\sigma$
confidence level. Finally, as we chose a uniform prior on $\log
c$, the confidence interval is not symmetric, resulting: $c^{2}=
0.005^{+0.007}_{-0.003}$.

\begin{figure}\centering
\includegraphics[scale=.9]{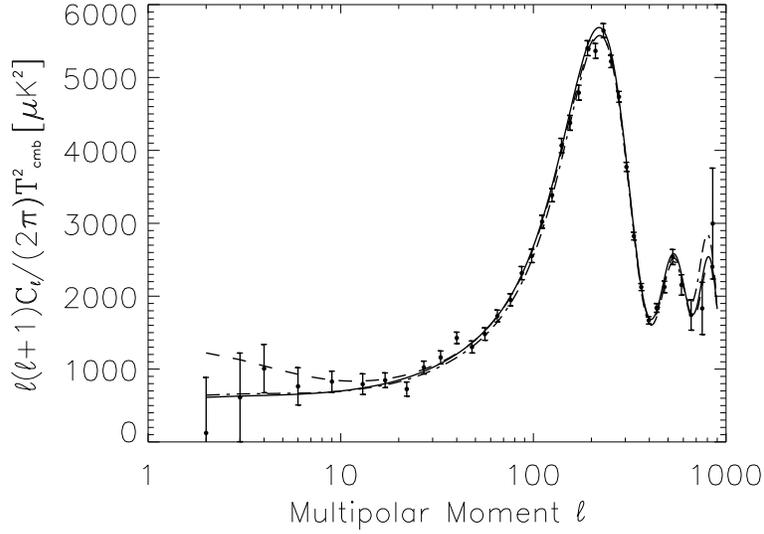}
\caption{Radiation Power Spectrum. The solid line is our best fit
model ($c^2=5\times 10^{-3}$, $\Omega_{x}=0.43$,
$\Omega_{b}=0.08$, $H_0=54\,km/s/Mpc$, $n_s=0.98$,$w=-0.99$).
Dashed line corresponds to the $\Lambda CDM$ concordance model and
dot-dashed line is $QCDM$ with parameters $\Omega_x=0.5$,
$\Omega_b = 0.07$, $H_{0}=60\,km/s/Mpc$, $w=-0.75$ and
$n_s=1.02$.} \label{fig:radpower}
\end{figure}

Our main result is that models with interaction are preferred over
non-interacting models, with the remarkable feature that they
require very different cosmological parameters than the
concordance model.
The range for $\Omega_{x}, H_{0}, \Omega_{b}$ and $w_x$ are not
directly comparable to those found in \cite{amendola} since the
interaction is different and we used different priors. Their
coupled model requires higher values of the Hubble constant when
the strength of the interaction increases, opposite to the
behavior found in Fig. \ref{fig:contour}.
Non-interacting models are compatible with
the data only at the 99.9\% confidence level. Our best fit model
($c^2=5\times 10^{-3}$, $\Omega_x=0.43$,
 $\Omega_b = 0.08$, $H_{0}=54\,km/s/Mpc$, $n_s=0.98$, $w=-0.99$)
has a $\chi^2=-2log(\cal{L}) = 974$ while the best fit for a
non-interacting model occurs at $\Omega_x=0.5$, $\Omega_b = 0.07$,
$H_{0}=60\,km/s/Mpc$, $w=-0.75$, $n_s=1.02$ and has $\chi^2 =
983$. The Bayesian Information Criteria defined as
$BIC=\chi^2+k\log N$ \cite{liddle}, that penalizes the inclusion
of additional parameters to describe data of small size (in this
case the number of independent data points is $N = 899$ and k, the
number of model parameters, is 5 in the non-IQ model, and 6 in the
model with interaction), gives $\Delta BIC = -2$ which can be
considered as positive evidence in favor of including this
additional parameter to describe the data. The $\Lambda$CDM
concordance model was deduced by fitting a different set of
parameters to WMAP data \cite{likeli} and the results are not
directly comparable to ours. For completeness, let us mention that
the concordance model with $\Omega_\Lambda=0.72$, $\Omega_b =
0.049$, $H_{0}= 68\,Km/s/Mpc$ and $n_s=0.97$ has a fit of $\chi^2
= 972$ is positively better than ours if the amplitude of the
matter power spectrum and the redshift of reionization are
included as parameters. If only the overall normalization of the
power spectrum is included, the fit is $\chi^2 = 990$. In this
case, the BIC would give $\Delta BIC = -9$ that must be taken as
strong indication that the interaction improves the fit.

The likelihood curves of Fig. \ref{fig:contour} seems to suggest
that our model is ruled out by observations since, for example,
luminosity distance estimates from high redshift supernovae
indicate that $\Omega_x\ge 0.6$ at the 95\% confidence level
\cite{hubble}. However, analysis of the temperature-luminosity
relation of distant clusters observed by XMM-Newton and Chandra
satellites appear to be consistent with $\Omega_{c}$ up to 0.8
\cite{cdm_high}. Moreover, with no priors on $\Omega_c$ and $w_x$,
(so phantom models are included in the analysis) values of the CDM
fraction as high as ours are consistent
 with SNIa data \cite{sn_review}.
Contours in Fig. \ref{fig:contour}c are rather parallel to the
$\Omega_x$-axis, while at the redshifts probed by supernovae
interacting models behave as non-interacting, and in the same
plane contours are parallel to the $c^2$-axis (see Fig.
\ref{fig:likeSNIQM}). In Fig. \ref{fig:contour}d we plot the
confidence intervals combining both WMAP and high redshift
supernovae data. Contours are shifted
to large values of dark energy, the reason being that WMAP data
constraints the coupling $c^2$ while the supernovae data is
insensitive to it. For illustrative purposes, in Fig.
\ref{fig:radpower} we compare our best fit model (solid line) with
the concordance model (dashed line) and the best quintessence
model with $c^2=0$ (dot-dashed). All models are rather smooth
compared with the rigging present in the data with the excess
$\chi^2$ coming from similar regions in $l$-space: $l\sim 120,
200, 350$. The discrepancy among the models is clearer at the
first acoustic peak. This rigging introduces a high degeneracy
among cosmological parameters since very different models fit the
data with similar $\chi^2$ per degree of freedom. When adding
other data sets, like the supernovae in Fig. \ref{fig:contour}c,
the fraction of dark energy increases dramatically. With respect
to the Hubble constant, the value is less certain since several
groups advocate values close to $60\,km/s/Mpc$ \cite{hubble_low},
and even as low as $50\,km/s/Mpc$ \cite{battistelli}.

There is a significant difference between the concordance
model  and our best fit model. Since the latter requires lower
Hubble constant and dark energy density, it generates a smaller
Integrated Sachs Wolfe effect, responsible for the rise of the
radiation power spectrum at $l\le 10$.  This is a generic
feature of this type of IQ models and, therefore, the low
amplitude of the measured quadrupole and octupole is less of a problem
in models with decaying dark energy than in the concordance model.

\section{Discussion}
We have shown that a model where the coincidence problem is solved
by the decay of the quintessence scalar field into cold dark
matter is fully compatible with the WMAP data. The best model,
$c^2\simeq 5 \times 10^{-3}$, fits the data significantly better
than models with no interaction. Our best fit model requires
cosmological parameters, in particular $\Omega_{x}$ and $H_{0}$,
that are very different from the concordance model. Our models are
highly degenerate in the $\Omega_{x}-c^{2}$ plane with contours
almost parallel to the $\Omega_{x}$ axis (see Fig.
\ref{fig:contour}) but including prior information from supernovae
data shifts this parameter close to the concordance model
$\Omega_{x} \simeq 0.68$. We wish to emphasize that the
non-interacting model ($c^{2} = 0$) is only compatible with WMAP data
at $3\sigma$ confidence level, but when the supernovae data are
included this is shifted to $2\sigma$ confidence level.

We have shown that the Bayesian Information Criteria, that
strongly disfavors increasing the parameter space to describe data
sets of $N\simeq 10^3$ points, provides positive evidence in favor
of the existence of interaction. Other IQ models have been
proposed \cite{amendola} and even if their results are not
directly comparable to ours, two interacting models, constructed
with different motivations, suggest a value of the dark energy
density smaller than in the concordance model, signals a need
to investigate this type of models. As we have discussed, the
quality of the fit and the cosmological parameters that can be
derived by fitting cosmological models  to observations depend on
the parameter space explored.

The fact that IQ models appear to be favored by current
observations suggests that dark energy and dark matter might not
be so different entities after all. This is in line with recent
ideas involving the Chaplygin gas. There, a single component plays
the dual role of cold dark matter (at early times) and vacuum
energy (at late times) and it interpolates between the  two as
expansion proceeds \cite{chgas}. The matter power spectrum could
also be an important test of IQ models. A preliminary study shows
that for the range of parameter space compatible with WMAP, the
effect of the scalar field decaying into CDM has little effect
\cite{progress}.

To summarize, the interacting cosmology model fits the WMAP data
significantly better than the $\Lambda$CDM model does, and in fact
alleviates the ISW effect at large angular scales, has no
coincidence problem and provides a unified picture of dark matter
and dark energy. It predicts lower values of Hubble constant, dark
energy density and higher baryon fraction. It is to be expected
that the next generation of CMB experiments \cite{planck} and
large scale surveys will enable us to constrain $c^{2}$ even
further and discriminate between the different variants.

\acknowledgments The authors wish to thank Alejandro Jakubi for
discussions and comments. This research was partially supported by
the Spanish Ministry of Science and Technology under Grants
BFM2003-06033, BFM2000-1322 and AYA2000-2465-E and the Junta de
Castilla y Le\'{o}n (project SA002/03).
%%%%%%%%%%%%%%%%%%%%%%%%%%%%%%%%%%%%%%%%%%%%%%%%%%%%%%%%%%%%%%%%%%%%%%%%%%%%%%%%


\begin{thebibliography}{99}
\bibitem{hubble}
S. Perlmutter {\it et al.}, Nature {\bf 391}, 51 (1998); A.G.
Riess {\it et al.}, Astrophys. J. {\bf 607}, 665 (2004).
\bibitem{wmap}
D.N. Spergel {\it et al.}, Astrophys. J. Supl. Ser. {\bf 148}, 175
(2003).
\bibitem{sdss}
M. Tegmark {\it et al.}, Phys. Rev. D {\bf 69}, 103501 (2004).
\bibitem{reviews}
S. Carroll, in {\it Measuring and Modelling the Universe},
Carnegie Observatory, Astrophysics Series, Vol. 2, edited by W.L.
Freedman (Cambridge University Press, Cambridge); T. Padmanbhan,
Phys. Reports, {\bf 380}, 235 (2003); J.A.S. Lima, Braz. J. Phys.
{\bf 34}, 194 (2004); V. Sahni, astro-ph/0403324; Proceedings of
the I.A.P. Conference {\em On the Nature of Dark Energ}, edited by
P. Brax {\it et al.} (Frontier Group, Paris, 2002); Proceedings of
the IVth Marseille Cosmology Conference {\em Where Cosmology and
Fundamental Physics Meet}, 23-26 June 2003, edited by V. Lebrun,
S. Basa and A. Mazure (Frontier Group, Paris, 2004).
\bibitem{peebles_rmph}
P.J.E. Peebles and B. Ratra, Rev. Mod. Phys. {\bf 75} 559 (2003).
\bibitem{boomerang}
P.D. Mauskopf {\it et al.}, Astrophys. Journal {\bf 536}, L59
(2000).
\bibitem{weinberg}
S. Weinberg, Rev. Mod. Phys. {\bf 61}, 1 (1989).
\bibitem{coincidence}
P.J. Steinhardt, in {\em Critical Problems in Physics},
edited by V.L. Fitch and D.R. Marlow
(Princeton University Press, Princeton, NJ, 1997).
\bibitem{lad}
L.P. Chimento, A.S. Jakubi and D. Pav\'on, Phys. Rev. D {\bf 62},
063508 (2000); {\it ibid.} {\bf 67}, 087302 (2003).
\bibitem{amendola}
L. Amendola, Phys. Rev. D {\bf 62}, 043511 (2000); L. Amendola
and D. Tocchini-Valentini, Phys. Rev. D {\bf 64}, 043509 (2001);
L. Amendola and D. Tocchini-Valentini, Phys. Rev. D {\bf 66}, 043528 (2002);
L. Amendola, C. Quercellini, D. Tocchini-Valentini and A. Pasqui,
Astrophys. J. {\bf 583}, L53(2003).
\bibitem{amendola0}
D. Tocchini-Valentini and L. Amendola, Phys. Rev. D {\bf 65}, 063508 (2002).
\bibitem{iqm0}
W. Zimdahl, D. Pav\'{o}n and L.P. Chimento,
Phys. Lett. B {\bf 521}, 133 (2001).
\bibitem{iqm}
L.P. Chimento, A.S. Jakubi, D. Pav\'{o}n and W. Zimdahl,
Phys. Rev. D {\bf 67}, 083513 (2003).
\bibitem{peebles}
G. Farrar and P.J.E, Peebles, Astrophys. J. {\bf 604}, 1 (2004).
\bibitem{hoffman}
B.M. Hoffman, astro-ph/9397359;  G. Huey and B. D. Wandelt,
astro-ph/0407196.
\bibitem {mangano}
G. Mangano, G. Miele and V. Pettorino, Modern. Phys.
Lett. A {\bf 18}, 831 (2003),
astro-ph/0212518.
\bibitem{rong}
Rong-Gen Cai and Anzhong Wang, hep-th/0411025.
\bibitem{peaks}
D. Pav\'{o}n, S. Sen and W. Zimdahl, J. Cosmology Astroparticle 05
(2004) 009.
\bibitem{scaling}
W. Zimdahl and D. Pav\'{o}n, Gen. Rel. Grav. {\bf 35}, 413 (2003).
\bibitem{gravity}
K. Hagiwara et al., Phys. Rev. D {\bf 66}, 010001 (2002).
\bibitem{liddle_scherrer}
A.R. Liddle and R.J. Scherrer, Phys. Rev. D {\bf 59}, 023509
(1999).
\bibitem{copeland}
E.J. Copeland, N.J. Nunes and F. Rosatti, Phys. Rev. D {\bf 62},
123503 (2000).
\bibitem{sahni_wang}
V. Sahni and L. Wang, Phys. Rev. D {\bf 62}, 103517 (2000).
\bibitem{riazuelo}
P. Brax, J. Martin and A. Riazuelo, Phys. Rev. D {\bf 62}, 103505 (2000).
\bibitem{one-loop}
M. Doran and J. J\"{a}ckel, Phys. Rev. D {\bf 66}, 043519 (2002).
\bibitem{bertschinger}
C.P. Ma and E. Bertschinger, Astrophys. J. {\bf 455}, 7 (1995).
\bibitem{caldwell}
R. Dave, R.R. Caldwell and P.J. Steinhardt, Phys. Rev. D {\bf 69},
023516 (2002).
\bibitem{cmbfast}
U. Seljak, M. Zaldarriaga,
Astrophysical Journal {\bf 469}, 437 (1996).
See http://www.cmbfast.org
\bibitem{likeli}
L. Verde {\it et al.}, Astrophys. J. Suppl. {\bf 148}, 195 (2003).
G. Hinshaw {\em et al.}, Astrophys. J. Suppl. {\bf 148}, 135
(2003); A. Kogut {\it et al.}, Astrophys. J. Suppl. {\bf 148}, 161
(2003).
\bibitem{olive}
K. Olive, TASI Lectures on Dark Matter, astro-ph/0301505.
\bibitem{cmb}
W. Hu and S. Dodelson,
Annu. Rev. Astron. and Astrophys. {\bf 40}, 171 (2002).
\bibitem{liddle}
G. Schwarz, Annals of Statistics {\bf 5}, 461 (1978); A.R. Liddle,
Month. Not. R. Astr. Soc. {\bf 351}, L49 (2004).
\bibitem{cdm_high}
S. C. Vauclair {\it et al.}, Astron. Astrophys. {\bf 412},  L37
(2003).
\bibitem{sn_review}
J. M. Virey {\it et al.}, Phys. Rev. D {\bf 70}, 121301 (2004).
\bibitem{hubble_low}
A. Blanchard {\it et al.}, Astron. Astrophys. {\bf 412}, 35
(2003); G. A. Tammann {\it et al.}, Astron. Astrophys. {\bf 404},
423 (2003).
\bibitem{battistelli} E. S. Battistelli {\it et al.},
Astrophys. J. {\bf 598}, L75 (2003).
\bibitem{chgas}
A. Kamenschik, U. Moschella and V. Pasquier, Phys. Lett. B {\bf
511}, 265 (2001); N. Bilic, G.B. Tupper and R.D. Viollier, Physics
Letters B {\bf 535}, 17 (2002); M. C. Bento, O. Bertolami, A. A.
Sen, Phys. Rev. D {\bf 66}, 043507 (2002).
\bibitem{progress} G. Olivares, F. Atrio-Barandela and
D. Pav\'{o}n, work in progress.
\bibitem{planck}
http://www.rssd.esa.int/index.php?project=PLANCK
\end{thebibliography}
\end{document}